\documentclass{PoS}
\pdfoutput=1
\usepackage{epsfig}
\usepackage{graphicx}

\newcommand{\Tr}{\mathrm{Tr}}

\title{`t Hooft model on the Lattice}

\ShortTitle{t Hooft model on the Lattice}

\author{ Margarita Garc\'{\i}a P\'erez$^{a}$, \speaker{Antonio
Gonz\'alez-Arroyo}
$^{a,b}$, Liam Keegan$^{c}$ and Masanori Okawa$^{d,e}$ \\
  $^a$ Instituto de F\'{\i}sica Te\'orica UAM-CSIC, Nicol\'as Cabrera 13-15, \\
  Universidad Aut\'onoma de Madrid, E-28049--Madrid, Spain \\
  $^b$ Departamento de F\'{\i}sica Te\'orica, C-15 \\
       Universidad Aut\'onoma de Madrid, E-28049--Madrid, Spain \\
  $^c$ Theoretical Physics Department, CERN, 1211 Geneva 23, Switzerland \\     
  $^d$ Graduate School of Science, Hiroshima University,\\
Higashi-Hiroshima, Hiroshima 739-8526, Japan \\
  $^e$ Core of Research for the Energetic Universe, Hiroshima University,\\
Higashi-Hiroshima, Hiroshima 739-8526, Japan \\

E-mail: \email{margarita.garcia@uam.es, antonio.gonzalez-arroyo@uam.es, liam.keegan@cern.ch, okawa@sci.hiroshima-u.ac.jp}
 }

\abstract{Lattice results are presented for the meson spectrum of
1+1 dimensional gauge theory at large $N$, using the Twisted
Eguchi-Kawai model. Comparison is made to the results obtained by `t
Hooft in the light cone  gauge.   
} 

\FullConference{34th annual International Symposium on Lattice Field Theory\\
		24-30 July 2016\\
		University of Southampton, UK}

\newcommand{\be}{\begin{equation}}
\newcommand{\ee}{\end{equation}}
\newcommand{\ba}{\begin{array}}
\newcommand{\ea}{\end{array}}
\newcommand{\baa}{\begin{array}}
\newcommand{\eaa}{\end{array}}
\newcommand{\bea}{\begin{eqnarray}}
\newcommand{\eea}{\end{eqnarray}}

\begin{document}
\section{Introduction}
`t Hooft model~\cite{thooft}-\cite{erice} is  just the 1+1 dimensional
version  of QCD in the large $N$ limit with quarks in the fundamental 
representation. Despite the huge expected differences with the 3+1 dimensional case (no
gluons, dimensionful constant, perturbative confinement, no angular
momentum, etc) the meson spectrum shows nice resemblances among both
cases (chiral limit, Regge behaviour, quark condensate~\cite{zhitnitsky}, etc). 
Let us briefly summarize the most salient features. The meson spectrum
consists in alternating even and odd parity states,  whose masses
$\mu_n$ asymptotically grow linearly with the integer index $n$:
\be
\mu_n^2 \sim n \pi^2 
\ee
The meson masses squared are expressed in units of  $m_0^2=g^2
N/\pi=\lambda/\pi$. Furthermore, the mass squared of the lowest meson state, which we
will call pion colloquially,  vanishes linearly with the bare quark mass as
follows: 
\be
\mu_0^2 = \frac{2\pi}{\sqrt{3}}m_q + \mathcal{O}(m_q^2)
\ee
This mass and those of the excited states can be obtained by solving a
1-dimensional integral equation. 
Recovering these results using Lattice Gauge Theories provides an interesting 
challenge. Some results~\cite{rebbi}  on 1+1 gauge theories have been obtained in the past 
by standard methods for low values of $N\le 5$.  However, no large $N$ 
extrapolation has been  attempted. The only quantitative study at large $N$ was done by Narayanan and
Neuberger~\cite{KNN} using the idea of reduced models~\cite{EK}.  Our
approach differs from this one by the use of the twisted reduced
model~\cite{TEK1}-\cite{TEK2}-\cite{TEK3} instead of the quenched
version~\cite{QEK}. We recently~\cite{meson,mesonlat15} proposed a method to compute
the meson spectrum in 3+1 dimensions in this context which applies
trivially also to the 1+1 dimensional case. It is certainly  a good
testing ground for our methodology since the meson masses in the
continuum are known from Ref.~\cite{thooft}.

\section{Methodology}
In this section we are going to briefly describe the main aspects of
our method. As shown in Ref.~\cite{EK} the expectation value of Wilson
loops in the large N limit becomes independent of the space-time volume. 
Hence it is possible to evaluate these loops by simulating a lattice 
model with a single point. This can be done either with the original
periodic Eguchi-Kawai prescription (EK), the quenched version (QEK)~\cite{QEK} or the
twisted version (TEK)~\cite{TEK1}-\cite{TEK2}. In 2D the conditions
necessary for  a proof of reduction are met in all cases.
Nevertheless, large $N$ corrections are much larger for EK than for
QEK and TEK. Paradoxically the calculation of rectangular 
Wilson loop expectation values  at large $N$ for the infinite volume 
setting is simpler and follows an exact area law~\cite{Gross:1980he} with a
string tension given by $-\log(1-1/(4b))$  where $b$ is the lattice
equivalent of the inverse `t Hooft coupling $\lambda$ corresponding to Wilson action. 

In this work we will be using the twisted version of the matrix model
whose partition function is given by 
\be
Z= \int dU_0\, dU_1 \ \exp\{b N z \Tr(U_0 U_1 U_0^\dagger U_1^\dagger
+\mathrm{h.c.})\}
\ee
where $b$ is the afore-mentioned lattice coupling ($b=1/\lambda_L$), the
matrices $U_\mu$ belong to SU($N$) and $z=e^{2 \pi i k/N}$ is an $N$-th
root of unity ($k$ is chosen co-prime with $N$). 

\begin{figure}[t]
   \begin{center}
    \includegraphics[height=4.8cm,  angle=0]{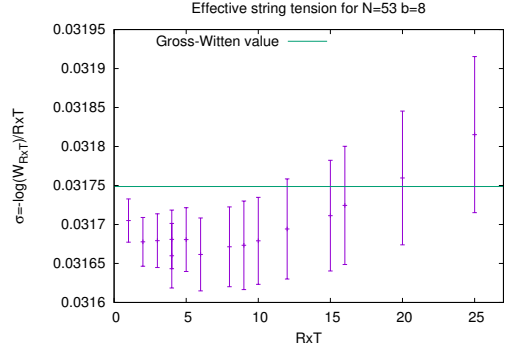}
        \end{center}
        \vspace{-0.7cm}
          \caption{Logarithm of the expectation value of the Wilson loop divided by the area.}
            \label{fig1}
	                                                     \end{figure}

In this theory rectangular Wilson loops can be constructed in terms of
the two matrices $U_0$ and $U_1$ as follows:
\be
W_{R\times T}= \frac{z^{RT}}{N} \langle \Tr(U^T_0 U _1^R U_0^{-T}
U_1^{-R}) \rangle
\ee
Volume reduction implies that in the large $N$ limit these expectation
values would coincide with those obtained for standard Wilson loops in
the infinite volume theory. Indeed, this can be tested. For example,
in Fig~\ref{fig1} minus the logarithm of the expectation value divided by
the area of the loop is plotted as a function of area for  several  
Wilson loops at $b=8$ and $N=53$. Agreement with the exact area law and
string tension evaluation is obtained to within a few  permille level. 

An important question when using the reduced model is what is the size
and nature of finite $N$ corrections. The perturbative
analysis~\cite{TEK2} reveals that part of the $1/N$ corrections has the same
form as that of a finite volume correction for a lattice volume of
size $N^2$. This idea is very helpful in determining the values of $N$
that one must choose to minimize the effect of finite $N$ errors. The 
conclusion is that the correlation lengths should be kept much smaller
than the effective length scale of our box, which is $N$.

Our main goal is to obtain the meson spectrum for this model. We will
follow the same procedure that we employed for the 4 dimensional
case~\cite{meson}. The idea is that, given that gauge fields are
effectively living in an $N\times N$ box, the same can be chosen for
the quark fields. Indeed, one can duplicate the size in the time
direction to avoid the need of using cosine hyperbolic instead of
simple exponentials when fitting the time dependence of correlators.  
Then, the analysis performed in Ref.~\cite{meson} implies that the 
meson correlators at zero spatial momentum are given by 
\be
C_{AB}(t) = \sum_{p_0} e^{i p_0 t} \langle \Tr( O_A
D^{-1}(p_0)O_B D^{-1}(0)) \rangle  
\ee
where $O_A$ and $O_B$ are operators defined at fixed time and are
meant to project onto meson states of the right quantum numbers. In
our case the only quantum number is parity. The symbol $D^{-1}(p_0)$
stands for the lattice quark propagator where the time dependence has
been Fourier transformed. In principle one is free to choose any
possible version of lattice fermions.
In this work we are using Wilson fermions, naive fermions and overlap
fermions. As a matter of fact, we can use our simple system to test the possible
advantages and disadvantages of each choice.  The spectra
can be different at fixed lattice spacing, but should agree
with `t Hooft values in the continuum limit. Here we will show that
nice results can be obtained for all cases, but we  defer a
full comparison for a future publication.

In any case, the propagator is obtained by inverting the corresponding
Dirac operator which is a $2N^2\times 2 N^2$ matrix. The covariant
derivative appearing in the operator has to be replaced by 
\be
D_\mu \longrightarrow U_\mu\otimes \tilde{\Gamma}_\mu - \mathbf{I}
\ee
where $U_\mu$ are the gauge matrices of the reduced model and the 
$N\times N$ matrices $\tilde{\Gamma}_\mu$ act on spatial coordinates. 
Its actual form is irrelevant because it is easy to see that 
they satisfy
\be
\tilde{\Gamma}_0 \tilde{\Gamma}_1 = z \tilde{\Gamma}_1 \tilde{\Gamma}_0
\ee
and this equation has a unique solution having
$\tilde{\Gamma}_\mu^N=\mathbf{I}$ modulo similarity
transformations. The latter are irrelevant since we are projecting
over zero spatial momentum.

Concerning the operators $O_A$ employed in our analysis we have
followed an analogous implementation to the one we used for the
four-dimensional case, reported by M. Okawa in this
conference~\cite{GAO}. One could choose ultralocal operators,
in which case they act only on the spinor indices. The 
unit matrix and the three Pauli matrices form a complete set. To
reduce the contribution of the excited states one can use operators
having a spatial extent. This is done by applying a {\em smearing
operator}  $D_s$ which adds a contribution one lattice spacing apart. 
This can be  iterated several times to give operators with different
smearing levels which can then be used in a variational determination
of the energy eigenstates. For details we refer the reader to our 
future publication.

We also  make a few comments about the numerical aspects of our
work. We have analysed data for various values of $N$: 31, 43 and 53
and various values of $b$ in the range $[3,8]$. 
In each case  we generated  1000 gauge field configurations using the 
over-relaxation method of Ref.~\cite{OR}. This step takes
minutes in a standard PC for these values of $N$. The matrix inversions
are performed with the source method. The procedure can be carried on
a PC or a small cluster. The total time employed depends on the number
of smearing levels considered.

As mentioned earlier in our analysis we use several different lattice
operators. As in our four dimensional studies we use  Wilson fermions.
This has  two major disadvantages. One is the existence of an additive
quark mass renormalization which forces to make a fit in the hopping
parameter to determine its critical value corresponding to zero bare
quark mass. On the other hand one expects lattice artifacts of order
$a$ rather than $a^2$. Our second option is to take naive fermions.
Both of the mentioned problems are solved at the expense of generating doubling.
However, in our case there are no dynamical fermions and  the doubling only 
affects valence quarks. This generates some inessential complications 
in extracting the meson masses.  Finally, we
also use massive overlap fermions, with the  conventions of Ref.~\cite{edwards}:
\be
D_{\mathrm{OV}}=\frac{1}{2}\left( 1+\mu + (1-\mu)
\gamma_5 \mathrm{sign}(\gamma_5 D_{\mathrm{WD}}(-m)) \right)
\ee
where $D_{\mathrm{WD}}(-m)$ is the Wilson Dirac operator with negative
mass $m$. The main disadvantage is that this takes longer to invert. However,
using several optimizations and  the multishift conjugate gradient method
to invert several masses simultaneously, the total time cost of using the
overlap  is only a factor 10 to 15 bigger than that for Wilson fermions.   

The final comment relates to the methodology used to extract the
masses. The most naive methodology would be perform a multi-exponential  
fit to the correlation function in time. A better alternative is to
use a variational analysis similar to the one employed in four
dimensions. The systematic errors of these methods arise from the
choice of range in which to perform the fits. This can be avoided if
one extracts the masses from a fit to the correlation function as a
function of $p_0$ as done in Ref.~\cite{KNN}. This, however, has
other types of systematic errors emerging from the parameterization of
this correlator. In any case, the final results of our study should show
the relative benefits of the different methods.  

\section{Results}

In this section we will present some of the preliminary results of our study,
separating the analysis between the different types of Dirac operators.

\begin{figure}[t]
 \begin{minipage}{0.5\hsize}
   \begin{center}
\vspace{-0.4cm}
\hspace*{-0.4cm}
      \includegraphics[width=75mm]{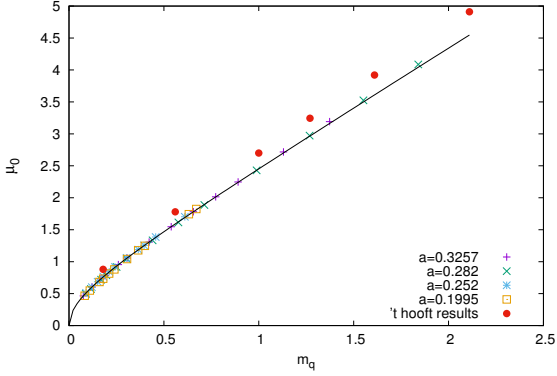}
        \end{center}
        \vspace{-0.7cm}
          \caption{Lowest meson mass for Wilson fermions}
            \label{fig2}
             \end{minipage}
              \begin{minipage}{0.5\hsize}
                \begin{center}
                   \includegraphics[width=75mm]{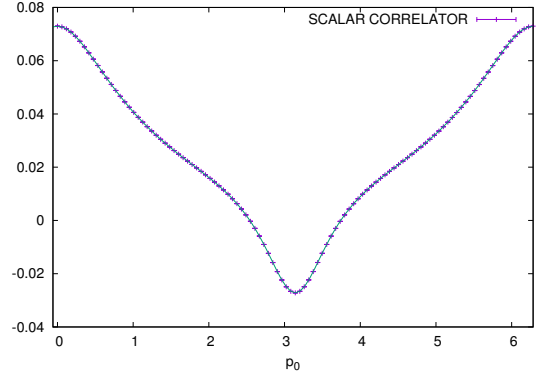}
                     \end{center}
                     \vspace{-0.7cm}
                       \caption{Correlator in momentum space for naive
                        fermions $N=53$ and $b=8$}
                         \label{fig3}
                          \end{minipage}
                          \end{figure}

\subsection{ Wilson fermions}
The procedure in this case is similar to the one that we have followed
in the four dimensional case. We generated 1000 configurations at $N=31$ 
and various values of $b$ (3, 4, 5 and 8), which correspond to a lattice
spacing of $a=1/\sqrt{\pi b}$ expressed in the mass unit mentioned
in the introduction. Then we computed the meson correlator for various
values of  the hopping parameter $\kappa$. This is converted to a
lattice bare  quark mass given by the formula 
\be
M_q=a\, m_q = \log(1+\frac{1}{2 \kappa}-\frac{1}{2 \kappa_c})
\ee
The value of $\kappa_c$ is obtained from a fit of the lowest meson mass
square to a linear plus quadratic function of $M_q$. Our results for
this meson mass are displayed in Fig.~\ref{fig2} as a function of $m_q$. 
They seem to scale quite well, but the values are smaller than the
corresponding values of Ref.~\cite{thooft}. At this stage it is hard to
know  what is the origin of the discrepancy. It could be due to finite
$N$ corrections, but most probably to a mismatch between the
definition of  quark mass on the lattice and the equivalent in
`t Hooft paper. Indeed, a rescaling of the lattice quark mass by a
 factor 0.9 makes the data fall on top of `t Hooft data points.

\begin{figure}[t]
 \begin{minipage}{0.5\hsize}
    \begin{center}
          \includegraphics[width=75mm]{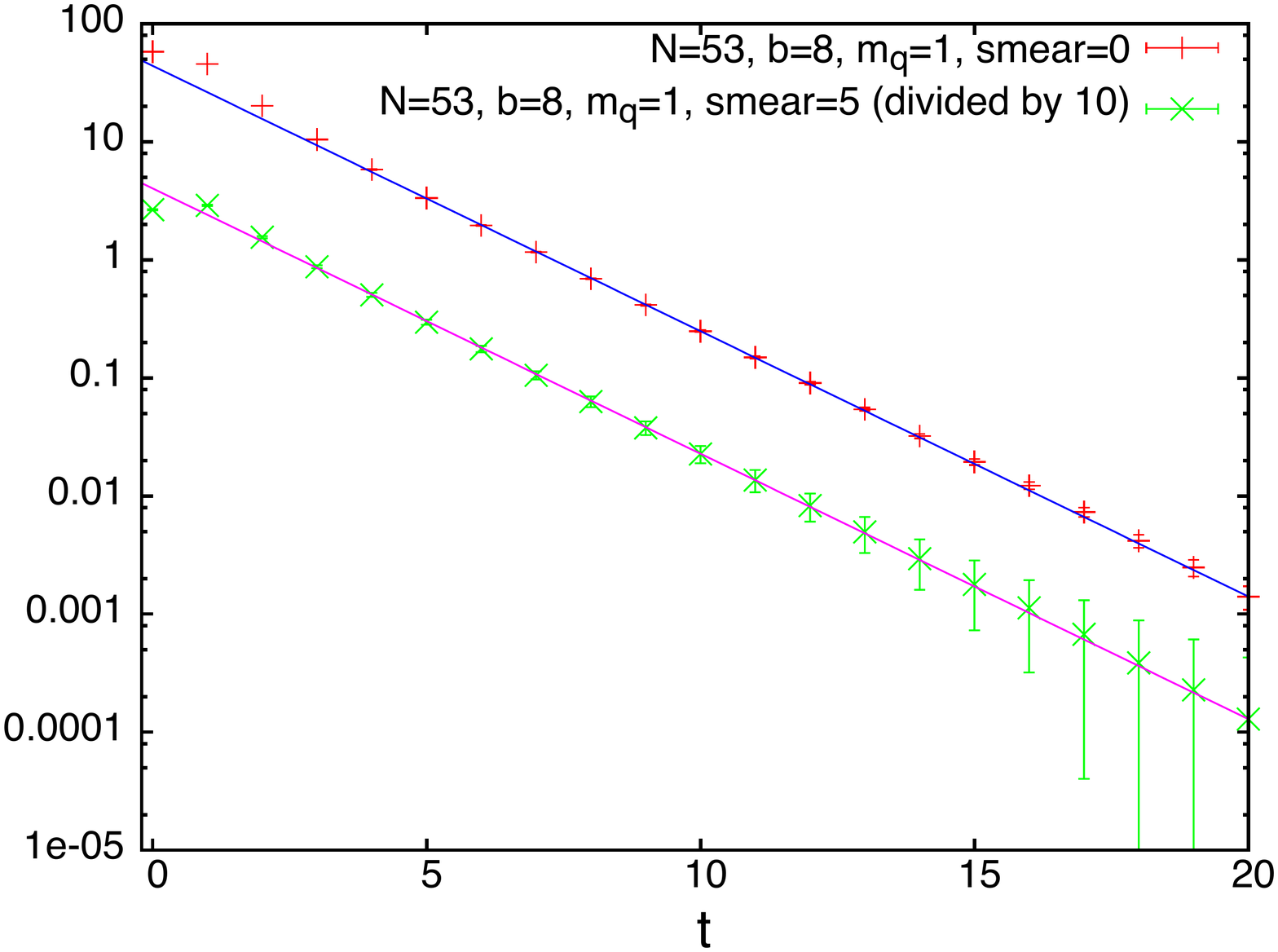}
                  \end{center}
                         \vspace{-0.8cm}
                            \caption{Correlators in position space for
                            $\quad \quad \quad \ $   overlap fermions.}
\label{fig4}
 \end{minipage}
 \begin{minipage}{0.5\hsize}
 \begin{center}
 \includegraphics[width=75mm]{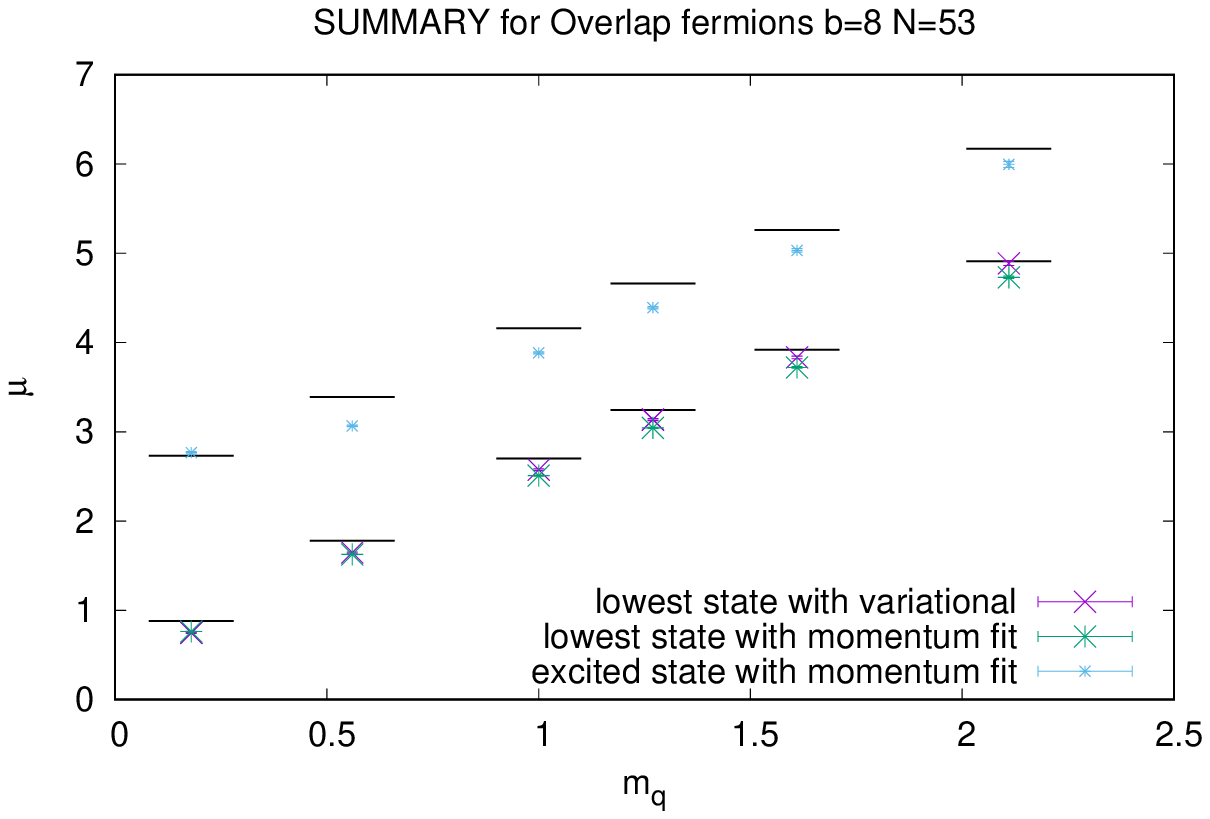}
 \end{center}
      \vspace{-0.7cm}
      \caption{Summary of meson masses for various quark masses for
      overlap fermions.}
    \label{fig5}
    \end{minipage}
    \end{figure}

\subsection{ Naive fermions}
The necessity to fit a value of critical $\kappa$ can be eliminated if
we study naive fermions, a special projection of which  gives
rise to staggered fermions. As a counterpart one has to deal with the
phenomenon of doubling. However, in this case this problem can be
easily handled. On one side there will be a double quasi-degeneracy of
all states. The degeneracy is only exact if $N$ is even. This doubling
of states demands a similar enlargement of the space of operators in a
variational analysis. In addition,
there is a symmetry of the quark propagator as follows:
\be
D^{-1}(p_0)\longrightarrow \gamma_1 D^{-1}(p_0+\pi)\gamma_1 
\ee
This implies that the meson propagator in $p_0$ space  contains
contributions of both parity even and odd states. For example, for the
$\bar{\psi}\psi$ case   we can then
parameterize its correlator as follows 
\be
C(p_0)= \sum_{i=0}^{M} \frac{c_{2i+1}}{\sin^2(p_0/2)+\sinh^2(\mu_{2i+1}
a/2)} -\sum_{j=0}^{M'} \frac{c_{2j}}{\cos^2(p_0/2)+\sinh^2(\mu_{2j}
a/2)}   
\ee
Very massive states give a constant contribution and only a few
individual masses can be extracted. Fig.~\ref{fig3} shows that a very
good description of the data ($\chi^2/\mathrm{ndof}=0.47$) can be obtained using this
parameterization with $M=M'=1$ (4 meson states) and a constant term mimicking
 the more massive states. A simultaneous fit to all
correlators and operators can be done in order to extract the lowest
lying masses. This can be done for all lattice fermion cases.

\subsection{Overlap fermions}
To conclude we give a brief snapshot of our results with the overlap
operator. In Fig.~\ref{fig4} we show the correlator in time for the
$\gamma_5=\gamma_1\gamma_2$ channel with and without smearing. The
exponential fall-off is  obvious after the first few values 
of $t$.  Smearing improves the agreement for smaller  values of $t$ 
but the errors become larger for higher $t$. In any case the two
operators give   compatible values for the mass. 

To conclude this presentation we display in Fig.~\ref{fig5} a summary
of our present results. The  two lowest-lying states are shown at various
values of the quark masses compared with the corresponding values as
given by `t Hooft shown as horizontal lines in the figure. The
qualitative agreement is quite clear, but the lattice values tend to
be slightly smaller. In any case  a full analysis of systematic errors
is still missing.

\section{Conclusions}
In this paper we have presented preliminary results of our analysis of
the meson spectra of 2 dimensional QCD in the large $N$ limit. Our
results are qualitatively in agreement with the continuum results of 
Ref.~\cite{thooft}. For a quantitative comparison an appropriate 
connection of the quark mass in the continuum and on the lattice is
necessary. The full results will be presented in a future publication. 
Our study serves as a testing ground for calculations of the meson
spectra for other theories with fermions in the adjoint or two-index
representation where the same techniques can be applied in the
analysis of results. 
\acknowledgments{
We acknowledge financial support from the 
grants FPA2012-31686, FPA2012-31880,  FPA2015-68541-P,
and the Spanish MINECO's ``Centro
de Excelencia Severo Ochoa'' Programme under grant
SEV-2012-0249. M. O. is supported by the Japanese MEXT grant No
26400249 and the MEXT program for promoting the enhancement of research
universities. Computations have been done with the clusters and PC's
at IFT.
}

\end{document}